\documentclass[aps,prx,twocolumn,nofootinbib,showpacs,superscriptaddress]{revtex4-1}

\usepackage{graphicx}

\usepackage[colorlinks=true,linkcolor=blue,citecolor=blue,urlcolor=blue]{hyperref}

\usepackage{amsmath}
\usepackage{multirow}
\usepackage{color}

\begin{document}

\title{Extending the predictive power of perturbative QCD using the principle of maximum conformality and the Bayesian analysis}

\author{Jian-Ming Shen}
\author{Zhi-Jian Zhou}
\affiliation{School of Physics and Electronics, Hunan Provincial Key Laboratory of High-Energy Scale Physics and Applications, Hunan University, Changsha 410082, People’s Republic of China}

\author{Sheng-Quan Wang}
\affiliation{Department of Physics, Guizhou Minzu University, Guiyang 550025, People’s Republic of China}

\author{Jiang Yan}
\author{Zhi-Fei Wu}
\author{Xing-Gang Wu}
\email{wuxg@cqu.edu.cn}
\affiliation{Department of Physics, Chongqing Key Laboratory for Strongly Coupled Physics, Chongqing University, Chongqing 401331, People’s Republic of China}

\author{Stanley J. Brodsky}
\email{sjbth@slac.stanford.edu}
\affiliation{SLAC National Accelerator Laboratory, Stanford University, Stanford, CA 94309, USA}

\begin{abstract}

In addition to the evaluation of high-order loop contributions, the precision and predictive power of perturbative QCD (pQCD) predictions depends on two important issues: (1) how to achieve a reliable, convergent fixed-order series, and (2) how to reliably estimate the contributions of unknown higher-order terms. The recursive use of renormalization group equation, together with the Principle of Maximum Conformality (PMC), eliminates the renormalization scheme-and-scale ambiguities of the conventional pQCD series. The result is a conformal, scale-invariant series of finite order which also satisfies all of the principles of the renormalization group. In this paper we propose a novel Bayesian-based approach to estimate the size of the unknown higher order contributions based on an optimized analysis of probability distributions. We show that by using the PMC conformal series, in combination with the Bayesian analysis, one can consistently achieve high degree of reliability estimates for the unknown high order terms. Thus the predictive power of pQCD can be greatly improved. We illustrate this procedure for two pQCD observables: $R_{e^+e^-}$ and $R_\tau$, which are each known up to four loops in pQCD. Numerical analyses confirm that by using the scale-independent and more convergent PMC conformal series, one can achieve reliable Bayesian probability estimates for the unknown higher-order contributions.



\end{abstract}

\maketitle

\section{Introduction}
\label{sec:intro}

Quantum chromodynamics (QCD) is the fundamental non-Abelian gauge theory of the strong interactions. Because of its property of asymptotic freedom, the QCD couplings between quarks and gluons become weak at short distances, allowing systematic perturbative calculations of physical observables involving large momentum transfer~\cite{Gross:1973id, Politzer:1973fx}. A physical observable must satisfy ``renormalization group invariance'' (RGI)~\cite{Petermann:1953wpa, GellMann:1954fq, Peterman:1978tb, Callan:1970yg, Symanzik:1970rt}; i.e., the infinite-order perturbative QCD (pQCD) approximant of a physical observable must be independent of artificially introduced parameters, such as the choice of the renormalization scheme or the renormalization scale $\mu_r$. A fixed-order pQCD prediction can violate RGI due to the mismatching of the scale of the perturbative coefficients with the corresponding scale of the strong coupling at each order. For example, invalid scheme-dependent predictions can be caused by an incorrect criteria for setting the renormalization scale; e.g., by simply choosing the scale to eliminate large logarithmic contributions. The error caused by the incorrect choice of the renormalization scale can be reduced to a certain degree by including enough higher-order terms and by the mutual cancellation of contributions from different orders. However, the complexity of high-order loop calculations in pQCD makes the available perturbative series terminate at a finite order, and thus the sought-after cancellations among different orders can fail. Clearly, as the precision of the experimental measurements is increased, it becomes critically important to eliminate theoretical uncertainties from the renormalization scale and scheme ambiguities and to also obtain reliable estimates of the contributions from unknown higher-order (UHO) terms.

Recall that in the case of high precision calculations in quantum electrodynamics (QED), the renormalization scale is chosen to sum all vacuum polarization contributions, e.g.
\begin{equation}
\alpha(q^2) = {\alpha(q^2_0)\over 1-\Pi (q^2, q^2_0)},
\end{equation}
where $\Pi(q^2, q^2_0) = (\Pi(q^2, 0)-\Pi(q^2_0, 0))/(1-\Pi(q^2_0, 0))$ sums all vacuum polarization contributions, both proper and improper, into the dressed photon propagator. This is the standard Gell Mann-Low renormalization scale-setting for perturbative QED series~\cite{GellMann:1954fq}.

Similarly, in non-Abelian QCD, the {\it Principle of Maximum Conformality} (PMC)~\cite{Brodsky:2011ig, Brodsky:2012rj, Brodsky:2011ta, Mojaza:2012mf, Brodsky:2013vpa} provides a rigorous method for obtaining a correct fixed-order pQCD series consistent with the principles of renormalization group~\cite{Brodsky:2012ms, Wu:2013ei, Wu:2014iba}. The evolution of the running QCD coupling is governed by the renormalization group equation (RGE),
\begin{equation}
\beta(\alpha_s)=\mu^2_r \frac{d \alpha_s(\mu_r)} {d \mu^2_r}=-\alpha_s^2(\mu_r) \sum_{i=0}^\infty \beta_i \alpha_s^{i}(\mu_r). \label{rgealpha}
\end{equation}
where the $\{\beta_i\}$-functions are now known up to five-loop level in the $\overline{\rm MS}$-scheme~\cite{Gross:1973ju, Politzer:1974fr, Caswell:1974gg, Tarasov:1980au, Larin:1993tp, vanRitbergen:1997va, Chetyrkin:2004mf, Czakon:2004bu, Baikov:2016tgj}. As in QCD, all $\beta$ terms are summed into the running coupling by the PMC. After PMC scale setting, the resulting pQCD series is then identical to the corresponding conformal theory with $\beta=0$. The PMC thus fixes the renormalization scale consistent with the RGE. It extends the Brodsky-Lepage-Mackenzie method~\cite{Brodsky:1982gc} for scale-setting in pQCD to all orders, and it reduces analytically to the standard scale-setting procedure of Gell-Mann and Low in the QED Abelian limit (small number of colors, $N_C \to 0$~\cite{Brodsky:1997jk}). The resulting relations between the predictions for different physical observables, called {\it commensurate scale relations}~\cite{Brodsky:1994eh, Huang:2020gic}, ensure that the PMC predictions are independent of the theorist's choice of the renormalization scheme. The PMC thus eliminates both renormalization scale and scheme ambiguities. As an important byproduct, because the RG-involved factorially divergent renormalon-like terms such as $n! \beta_0^n \alpha_s^n$~\cite{Beneke:1994qe, Neubert:1994vb, Beneke:1998ui} are eliminated, the convergence of the PMC perturbative series is automatically improved. In contrast, if one guesses the renormalization scale such as choosing it to match the factorization scale, one will obtain incorrect, scheme-dependent, factorially divergent results for the pQCD approximant, as well as violating the analytic $N_C\to 0$ Abelian limit. Such an {\it ad hoc} procedure will also contradict the unification of the electroweak and strong interactions in a grand unified theory.

In practice, it has been conventional to take $\mu_r$ as the typical momentum flow ($Q$) of the process in order to obtain the central value of the pQCD series and to then vary $\mu_r$ within a certain range, such as $[Q/2, 2Q]$, as a measure of a combined effect of scale uncertainties and the contributions from uncalculated higher-order (UHO) terms. The shortcomings of this {\it ad hoc } treatment are apparent: each term in the perturbative series is scale-dependent, and thus the prediction will not satisfy the requirement of RGI. Furthermore, an estimate of the UHO contributions cannot be characterized in a statistically meaningful way; one can only obtain information for the $\beta$-dependent terms in the uncalculated higher-order terms which control the running of $\alpha_s$, and there are no constraints on the contribution from the higher-order conformal $\beta$-independent terms.

Since the exact pQCD result is unknown, it would be helpful to quantify the UHO's contribution in terms of a probability distribution. The {\it Bayesian analysis} is a powerful method to construct probability distributions in which Bayes' theorem is used to iteratively update the probability as new information becomes available. In this article, we will show how one can apply the Bayesian analysis to predict the uncertainty of the UHO contributions as a weighted probability distribution. This idea was pioneered by Cacciari and Houdeau \cite{Cacciari:2011ze}, and has been developed more recently in Refs.\cite{Bagnaschi:2014wea, Bonvini:2020xeo, Duhr:2021mfd}. As illustrations of the power of this method, we will apply the Bayesian analysis to estimate the UHO contributions to several hadronic QCD observables.

Previous applications of Bayesian-based approach have been based on highly scale-dependent pQCD series. As discussed above, it clearly is important to instead use a renormalization scale-invariant series as the basis in order to show the predictive power of the Bayesian-based approach. In the paper, we shall adopt the PMC scale-invariant conformal series as the starting point for estimating the magnitude of unknown higher order contributions using the Bayesian-based approach.

The remaining parts of the paper are organized as follows: In Sect.\ref{sec:bayes} we show how the Bayesian analysis can be applied to estimate the contributions of the UHO terms. In Sect.\ref{sec:pmc}, we will give a mini-review of how high precision predictions can be achieved by using the PMC single scale-setting approach (PMCs). In Sect.\ref{sec:numer}, we will apply the PMCs and the Bayesian-based approach to give predictions with constrained high order uncertainties for two observables, $R_{e^+ e^-}$ and $R_{\tau}$, Sect.\ref{sec:summary} is reserved for a summary. For convenience and as a useful reference, we provide a general introduction to probability and the Bayesian analysis~\cite{Workman:2022ynf}, together with a useful glossary of the terminology in the Appendix.

\section{The Bayesian-based approach}
\label{sec:bayes}

In this section we will show how one can apply a Bayesian-based approach in order to give a realistic estimate of the size of the unknown higher order pQCD contributions to predictions for physical observables. We shall show that by using the PMC conformal series, in combination with the Bayesian analysis of probability distributions, one can consistently achieve a high degree of reliability estimate for the UHO-terms. Thus the predictive power of pQCD can be greatly improved.

We will explain the Bayesian-based approach by applying it to the series of a perturbatively calculable physical observable ($\rho$). If the perturbative approximant of the physical observable starts at order $\mathcal{O}(\alpha_s^l)$ and stops at the $k_{\rm th}$ order $\mathcal{O}(\alpha_s^k)$, one has
\begin{eqnarray}
\label{eq:rhok}
\rho_{k} = \sum_{i=l}^k c_i \alpha_s^i,
\end{eqnarray}
which represents the partial sum consisting of the first several terms in the perturbative expansion. $c_i$ are the coefficients of the perturbative expansion.

For conventional pQCD series, the limit, $k\to \infty$, does not exist, as perturbative expansions are divergent \cite{Dyson:1952tj, tHooft:1977xjm}. The typical divergent contributions are renormalons (see e.g. Ref.\cite{Beneke:1998ui}).
The divergent nature of the pQCD series is related to the fact that $\rho$ is a non-analytic function of the coupling $\alpha_s$ in $\alpha_s=0$.
The conventional pQCD series are believed to be asymptotic expansions of the physical observable.

The asymptotic nature of the divergent perturbative expansion implies that up to some order $N$ adding terms to the expansion improves the accuracy of the prediction, but beyond $N$ the divergent contributions to the series dominate and the sum explodes. The truncated expansion at order $N$ gives the optimally accurate approximation available for the observable ($\rho$), and represents the {\it optimal truncation} of the asymptotic series.

With these statements, following we will give a brief introduction on how to apply the Bayesian-based approach to a fixed-order pQCD series, and estimate the size of its unknown higher orders in terms of the properties of a probability distribution. Other applications and developments of the Bayesian-based method can be found in Refs.\cite{Cacciari:2011ze, Bagnaschi:2014wea, Bonvini:2020xeo, Duhr:2021mfd}.

\subsection{Basic definitions and assumptions}

Consider a generic measure of ``credibility'', applicable to any possible perturbative series such as Eq.(\ref{eq:rhok}), varied over the space of a set of {\it priori } unknown perturbative coefficients $c_l,c_{l+1},\cdots$. These coefficients are regarded as random variables in Bayesian statistics. One can define a {\it probability density function} (p.d.f.), $f(c_{l},c_{l+1},\cdots)$, which satisfies the following normalization condition
\begin{eqnarray}
\label{eq:normalization}
\int f(c_l,c_{l+1},\cdots)\ {\rm d} c_{l}\ {\rm d} c_{l+1} \cdots = 1,
\end{eqnarray}
and the parameters can be marginalized according to
\begin{eqnarray}
\label{eq:marginalize}
f(c_l,\cdots,c_k)=\int f(c_l,\cdots,c_k,c_{k+1},\cdots)\ {\rm d}c_{k+1} \cdots .
\end{eqnarray}
If not specified, here and following, the ranges of integration for the variables are all from $-\infty$ to $+\infty$. The conditional p.d.f. of a generic (uncalculated) coefficient $c_n$ with given coefficients $c_l,\dots,c_k$, is then by definition,
\begin{eqnarray}
\label{eq:conditionalpdf}
f_c(c_n|c_l,\dots,c_k)=\frac{f(c_l,\dots,c_k,c_{n})}{f(c_l,\dots,c_k)}, \;\; (n>k).
\end{eqnarray}

A key point of the Bayesian-based approach is to make the reasonable assumption that all the coefficients $c_i$ ($i=l,l+1,\cdots$) are finite and bounded by the absolute value of a common number $\bar{c}$ (${\bar c}>0$)~\cite{Cacciari:2011ze}, namely
\begin{equation}
\label{eq:cbar}
\left|c_i\right|\leq \bar{c}, \;\;\; \forall \; i.
\end{equation}
If none of the coefficients have been calculated, one can only say that $\bar c$ is a positive real number where its order of magnitude is {\it priori} unknown. If the first several coefficients such as $c_l,\cdots,c_k$ have been calculated, one may use them to give an estimate of ${\bar c}$, which in turn restricts the possible values for the unknown coefficient $c_n$ ($n>k$). The value of $\bar{c}$ is thus a (hidden) parameter which will disappear (through marginalization) in the final results. The set of uncertain variables that defines the space is thus the set constituted by the parameter ${\bar c}$ and all of the coefficients $c_l,c_{l+1},\cdots$. Three reasonable hypotheses then follow from the above assumption (\ref{eq:cbar}); i.e.
\begin{itemize}
\item The order of magnitude of $\bar{c}$ is equally probable for all values. This can be encoded by defining a p.d.f. for $\ln\bar{c}$, denoted by $g(\ln\bar c)$, as the limit of a flat distribution within the region of $-|\ln\epsilon|\leq\ln{\bar c}\leq|\ln\epsilon|$, where $\epsilon$ is a small parameter tends to $0$,
\begin{eqnarray}
\label{eq:epsdep0} g(\ln\bar c)=\frac{1}{2|\!\ln\epsilon|}\ \theta(|\ln\epsilon|-|\ln\bar c|).
\end{eqnarray}
Equivalently, a p.d.f. for ${\bar c}$, which is denoted by $g_0(\bar c)$, satisfies
\begin{eqnarray}
\label{eq:cbarpdf} g_0(\bar c)=\frac{1}{2|\!\ln\epsilon|}\frac{1}{\bar c}\ \theta\left(\frac{1}{\epsilon}-{\bar c}\right) \theta({\bar c}-\epsilon),
\end{eqnarray}
where $\theta(x)$ is the Heaviside step function. In practice we will perform all calculations (both analytical and numerical) with $\epsilon\neq 0$, and take the limit $\epsilon\rightarrow 0$ for the final result.

\item The conditional p.d.f. of an unknown coefficient $c_i$ given ${\bar c}$, which is denoted by $h_0(c_i|{\bar c})$, is assumed in the form of a uniform distribution, i.e.,
\begin{eqnarray}
\label{eq:cipdf} h_0(c_i|\bar c) = \frac{1}{2\bar c} \theta({\bar c}-|c_i|),
\end{eqnarray}
which implies that the condition (\ref{eq:cbar}) must be strictly satisfied. The p.d.f. $h_0(c_i|\bar c)$ will act as the \emph{likelihood function} for $\bar c$ in later calculations.

\item All the coefficients $c_i$ ($i=l,l+1,\cdots$) are mutually \emph{independent}, with the exception for the common bound, i.e. $\left|c_i\right|\leq \bar{c}$, which implies the joint conditional p.d.f., denoted by $h(c_i,c_j|\bar c)$,
\begin{eqnarray}
\label{eq:jointpdf} h(c_i,c_j|\bar c)=h_0(c_i|\bar c)h_0(c_j|\bar c), \;\; \forall \;\; i\neq j.
\end{eqnarray}
\end{itemize}
The hypotheses (\ref{eq:cbarpdf}), (\ref{eq:cipdf}) and (\ref{eq:jointpdf}) completely define the \emph{credibility measure} over the whole space of a priori uncertain variables $\{{\bar c},c_l,c_{l+1},\cdots\}$. They then define every possible inherited measure on a subspace associated with the pQCD approximate of a physical observable whose first several coefficients are known.

One may question the reasonability of the original assumption (\ref{eq:cbar}) due to the fact that the full pQCD series is divergent. However, in practice, of all the unknown higher orders, we shall concentrate on the terms before the optimal truncation. For all the terms before the optimal truncation $N$, it is reasonable to give a finite common boundary, $\bar c$, for their coefficients. For definiteness, we modify the assumption (\ref{eq:cbar}) as,
\begin{equation}
\label{eq:cbar1}
\left|c_i\right|\leq \bar{c}, \;\;\; \forall \; i \leq N \;.
\end{equation}
This modification will not change the above three hypotheses (\ref{eq:cbarpdf}, \ref{eq:cipdf}, \ref{eq:jointpdf}).

\subsection{Bayesian analysis}

In this subsection, we calculate the conditional p.d.f. of a generic (uncalculated) coefficient $c_n$ ($n>k$) with given coefficients $c_l,\cdots,c_k$, denoted as $f_c(c_{n}|c_l,\dots,c_k)$, based on the Bayes' theorem.

Schematically, we first reformulate the conditional p.d.f. $f_c(c_{n}|c_l,\dots,c_k)$ as,
\begin{eqnarray}
\label{eq:conditionalpdf1}
f_c(c_n|c_l,\cdots,c_k)=\int h_0(c_n|\bar c)f_{\bar c}(\bar c|c_l,\cdots,c_k) {\rm d}{\bar c},\;\;
\end{eqnarray}
where $f_{\bar c}({\bar c}|c_l,\cdots,c_k)$ is the conditional p.d.f. of ${\bar c}$ given $c_l,\cdots,c_k$. Applying Bayes' theorem, we have
\begin{eqnarray}
\label{eq:bayes}
f_{\bar c}(\bar c|c_l,\cdots,c_k)=\frac{h(c_l,\cdots,c_k|\bar c)g_0(\bar c)}{\int h(c_l,\cdots,c_k|\bar c)g_0(\bar c) {\rm d}{\bar c}}\;,
\end{eqnarray}
where $h(c_l,\cdots,c_k|\bar c)=\prod_{i=l}^k h_0(c_i|\bar c)$ according to (\ref{eq:jointpdf}) is the \emph{likelihood function} for $\bar{c}$. Inserting the Bayes' formula (\ref{eq:bayes}) and the factorization property (\ref{eq:jointpdf}) into (\ref{eq:conditionalpdf1}), and taking the limit $\epsilon\to 0$ for the final result, one obtains
\begin{eqnarray}
\label{eq:cnpdf}
&&\hspace{-3mm} f_c(c_n|c_l,\dots,c_k) = \lim_{\epsilon\to 0} \frac{\int h_0(c_n|\bar c)\prod_{i=l}^k h_0(c_i|\bar c)g_0(\bar c){\rm d}{\bar c}}{\int \prod_{i=l}^k h_0(c_i|\bar c)g_0(\bar c) {\rm d}{\bar c}} \nonumber\\
&&\hspace{+22mm} = \frac{1}{2}\frac{n_c}{n_c+1}\ \frac{\bar{c}_{(k)}^{n_c}}{(\max\{|c_n|,\bar{c}_{(k)}\})^{n_c+1}} \nonumber\\
&&\hspace{+22mm} =\left\{ \begin{array}{l l}
\frac{n_c}{2(n_c+1)\bar{c}_{(k)}}, & \; |c_n| \leq \bar{c}_{(k)} \vspace{2mm}\\
\frac{n_c{\bar{c}_{(k)}^{n_c}}}{2(n_c+1)|c_n|^{n_c+1}}, & \; |c_n| > \bar{c}_{(k)} \\
\end{array}
\right. .
\end{eqnarray}
where $\bar{c}_{(k)}=\max\{|c_l|,\cdots,|c_k|\}$, and $n_c=k-l+1$ represents the number of known perturbative coefficients, $c_l,\cdots,c_k$. It is easy to confirm the normalization condition, $\int^{\infty}_{-\infty}f_c(c_n|c_l,\dots,c_k){\rm d}c_n=1$. Equation (\ref{eq:cnpdf}) indicates the conditional p.d.f. $f_c(c_n|c_l,\dots,c_k)$ depends on the entire set of the calculated coefficients via $\bar{c}_{(k)} =\max\{|c_l|,\cdots,|c_k|\}$. The existence of such a probability density distribution within the uncertainty interval represents the main difference with other approaches, such as the conventional scale variation approach, which only gives an interval without a probabilistic interpretation. Equation (\ref{eq:cnpdf}) also shows a symmetric probability distribution for negative and positive $c_n$, predicts a uniform probability density in the interval $[-\bar{c}_{(k)},\bar{c}_{(k)}]$ and decreases monotonically from $\bar{c}_{(k)}$ to infinity. The knowledge of probability density $f_c(c_n|c_l,\dots,c_k)$ allows one to calculate the degree-of-belief (DoB, also called ``Bayesian probability'' or ``subjective probability'' or ``credibility'') that the value of $c_n$ is constrained within some interval. The smallest credible interval (CI) of fixed $p\%$ DoB for $c_n$ ($n>k$) turns out to be centered at zero, and thus we denote it by $[-c_n^{(p)},c_n^{(p)}]$. It is defined implicitly by
\begin{eqnarray}
\label{eq:cnDoB} 
p\% = \int_{-c_n^{(p)}}^{c_n^{(p)}} f_c(c_n |c_l,\dots,c_k)\ {\rm d} c_n,
\end{eqnarray}
and further by using the analytical expression in Eq.(\ref{eq:cnpdf}), we obtain
\begin{eqnarray}
\label{eq:cnp}
\hspace{-5mm}
c_n^{(p)}=
\left\{
\begin{array}{l l}
\bar{c}_{(k)} \frac{n_c+1}{n_c} p\%, & \; p\% \leq \frac{n_c}{n_c+1} \vspace{2mm}\\
\bar{c}_{(k)} \left[(n_c+1)(1-p\%)\right]^{-\frac{1}{n_c}}, & \; p\% > \frac{n_c}{n_c+1} \\
\end{array}
\right..
\end{eqnarray}

With the help of Eq.(\ref{eq:cnpdf}), one can then derive the conditional p.d.f. for the uncalculated higher order term $\delta_n=c_n\alpha_s^n$, ($n>k$), and the smallest $p\%$-CI for $\delta_{n}$, namely, $[-c_n^{(p)}\alpha_s^n,c_n^{(p)}\alpha_s^n]$. For the next UHO, i.e. $n=k+1$, the conditional p.d.f. of $\delta_{k+1}$ given coefficients $c_l,\dots,c_k$, denoted by $f_\delta(\delta_{k+1}|c_l,\cdots,c_k)$, reads,
\begin{widetext}
\begin{eqnarray}\label{eq:deltapdf}
f_\delta(\delta_{k+1}|c_l,\dots,c_k) = \left(\frac{n_c}{n_c+1}\right)\frac{1}{2\alpha_s^{k+1} \bar{c}_{(k)}}
\left\{
\begin{array}{ll}
1, & |\delta_{k+1}|\leq \alpha_s^{k+1}\bar{c}_{(k)}\\[8pt]
\left(\frac{\alpha_s^{k+1}\bar{c}_{(k)}}{|\delta_{k+1}|}\right)^{n_c+1}, & |\delta_{k+1}|>\alpha_s^{k+1}\bar{c}_{(k)}
\end{array}
\right. ,
\end{eqnarray}
\end{widetext}
Equation (\ref{eq:deltapdf}) indicates an important characteristic of the posterior distribution: a central plateau with power suppressed tails. The distributions for $\rho_{k+1}$ and $\delta_{k+1}$ are the same, up to a trivial shift given by the perturbative result (\ref{eq:rhok}). Thus the conditional p.d.f. of $\rho_{k+1}$ for given coefficients $c_l,\dots,c_k$, denoted by $f_\rho(\rho_{k+1}|c_l,\cdots,c_k)$, can be obtain directly,
\begin{widetext}
\begin{eqnarray}
\label{eq:rhopdf}
f_\rho(\rho_{k+1}|c_l,\cdots,c_k) = \left(\frac{n_c}{n_c+1}\right)\frac{1}{2\alpha_s^{k+1} \bar{c}_{(k)}}
\left\{
\begin{array}{ll}
1, & |\rho_{k+1}-\rho_{k}|\leq \alpha_s^{k+1}\bar{c}_{(k)}\\[8pt]
\left(\frac{\alpha_s^{k+1}\bar{c}_{(k)}}{|\rho_{k+1}-\rho_{k}|}\right)^{n_c+1}, & |\rho_{k+1}-\rho_{k}|>\alpha_s^{k+1}\bar{c}_{(k)}
\end{array}
\right. \, .
\end{eqnarray}
\end{widetext}
We can also estimate more UHOs of the perturbative series (\ref{eq:rhok}), e.g. the sum from the next UHO to the optimal truncation, $\Delta_k = \sum_{i=k+1}^{N} c_i \alpha_s^i$. The detail p.d.f. formulas of $\Delta_k$ are given in the appendix. In this work we shall concentrate on estimating the next UHO, $c_{k+1}$, for given coefficients $c_l,\dots,c_k$.

In the case of the conventional pQCD series, where the coefficients 
$\{c_l,c_{l+1},\cdots,c_k\}$ are renormalization scale dependent, the smallest CI, e.g. $[-c_n^{(p)},c_n^{(p)}]$, for the DoB of the coefficient $c_n$ under the fixed probability $p\%$ is also scale dependent. In order to achieve the goal of the Bayesian Optimization suggested by Refs.\cite{BO1, BO2}, i.e., to achieve the optimal smallest CI for the UHO by using the least possible number of given terms, it is clearly better to use a perturbative series with scale-invariant coefficients; i.e.,
\begin{eqnarray}
\frac{\partial}{\partial \mu_r^2}c_i = 0, \;\; \forall \;\; i \;.
\end{eqnarray}
For a general pQCD approximant $\rho_k$, such as Eq.(\ref{eq:rhok}), it is easy to confirm that
\begin{eqnarray}
\mu_r^2\frac{\partial \rho_{k}}{\partial \mu_r^2}\bigg|_{c_i} = -\beta(\alpha_s)\frac{\partial}{\partial \alpha_s}\rho_{k},
\end{eqnarray}
where the subscript $c_i$ means the partial derivative is done with respect to the perturbative coefficients only. It shows that if a perturbative series satisfies $\beta(\alpha_s)=0$, its coefficients will be scale-invariant. The PMC series satisfies this requirement by definition, and thus is well matched to achieve the goal of Bayesian Optimization. Our numerical results given in the following Sect.\ref{sec:numer} shall confirm this point.

\subsection{Consistent estimate for the contribution of unknown high order pQCD contributions}

One can calculate the expectation value and the standard deviation for $c_n$, $\delta_{k+1}$, and $\rho_{k+1}$ according to the p.d.f.s (\ref{eq:cnpdf}), (\ref{eq:deltapdf}) and (\ref{eq:rhopdf}), respectively. The expectation value and the standard deviation are the essential parameters. In the following, we shall adopt the determination of $\rho_{k+1}$ as an illustration.

It is conventional to estimate the central value of $\rho_{k+1}$ as its expectation value $E(\rho_{k+1})$ and estimate the theoretical uncertainty of $\rho_{k+1}$ as its standard deviation, $\sigma_{k+1}$. The expectation value $E(\rho_{k+1})$ can be related to the expectation value of $\delta_{k+1}$, i.e. $E(\rho_{k+1})=E(\delta_{k+1})+\rho_k$. For the present prior distribution, $E(\delta_{k+1})=0$, due to the fact that the symmetric probability distribution (\ref{eq:deltapdf}) is centered at zero. To predict the next UHO, $\delta_{k+1}$, of $\rho_{k}$ consistently, it is useful to define a critical DoB, $p_c\%$, which equals to the least value of $p\%$ that satisfies the following equations,
\begin{eqnarray}
\hspace{-5mm} \rho_{i-1}+c_{i}^{(p)}\alpha_s^{i}\geq \rho_{i}+c_{i+1}^{(p)}\alpha_s^{i+1}, \; (i=l+1,\cdots,k), \\
\hspace{-5mm} \rho_{i-1}-c_{i}^{(p)}\alpha_s^{i}\leq \rho_{i}-c_{i+1}^{(p)}\alpha_s^{i+1}, \; (i=l+1,\cdots,k).
\end{eqnarray}
Thus, for any $p\geq p_c$, the error bars determined by the $p\%$-CIs provide consistent estimates for the next UHO, i.e. the smallest $p\%$-CIs ($p\geq p_c$) of $\rho_{i+1}$ predicted from $\rho_{i}$ are well within the smallest $p\%$-CIs of the one-order lower $\rho_{i}$ predicted from $\rho_{i-1}$, ($i=l+1,l+2,\cdots,k$). The value of $p_c$ is nondecreasing when $k$ increases. In practice, in order to obtain a consistent and high DoB estimation, we will adopt the smallest $p_s\%$-CI; i.e.
\begin{eqnarray}
[E(\rho_{k+1})-c_{k+1}^{(p_s)}\alpha_s^{k+1},E(\rho_{k+1})+c_{k+1}^{(p_s)}\alpha_s^{k+1}],
\end{eqnarray}
as the final estimate for $\rho_{k+1}$, where $p_s={\rm max}\{p_c,p_\sigma\}$. Here $p_\sigma\%$ represents the DoB for the $1\sigma$-interval, and $\rho_{k+1}\in[E(\rho_{k+1})-\sigma_{k+1},E(\rho_{k+1})+\sigma_{k+1}]$.

\section{The Principle of Maximum Conformality}
\label{sec:pmc}

The PMC was originally introduced as a multi-scale-setting approach (PMCm)~\cite{Brodsky:2011ta, Brodsky:2012rj, Mojaza:2012mf, Brodsky:2013vpa}, in which distinct PMC scales at each order are systematically determined in order to absorb specific categories of $\{\beta_i\}$-terms into the corresponding running coupling $\alpha_s$ at different orders. Since the same type of $\{\beta_i\}$-terms emerge at different orders, the PMC scales at each order can be expressed in perturbative form. The PMCm has two kinds of residual scale dependence due to the unknown perturbative terms~\cite{Zheng:2013uja}; i.e., the last terms of the PMC scales are unknown (\textit{first kind of residual scale dependence}), and the last terms in the pQCD approximant are not fixed since its PMC scale cannot be determined (\textit{second kind of residual scale dependence}). Detailed discussions of the residual scale dependence can be found in the reviews~\cite{Wu:2019mky, Huang:2021hzr}. The PMC single-scale-setting approach (PMCs)~\cite{Shen:2017pdu} has been recently suggested in order to suppress the residual scale dependence and to make the scale-setting procedures much simpler. The PMCs procedure determines a single overall effective $\alpha_s$ with the help of RGE; the resulting PMC renormalization scale represents the overall effective momentum flow of the process. The PMCs is equivalent to PMCm in the sense of perturbative theory, and the PMCs prediction is also free of renormalization scale-and-scheme ambiguities up to any fixed order~\cite{Wu:2018cmb}. The PMCs is also equivalent to the very recently suggested single-scale-setting method~\cite{Yan:2022foz}, which follows the idea of ``Intrinsic Conformality''~\cite{DiGiustino:2020fbk}. By using the PMCs, the \textit{first kind of residual scale dependence} will be greatly suppressed due to its $\alpha_s$-power suppression and the exponential suppression; the overall PMC scale has the same precision for all orders, and thus the \textit{second kind of residual scale dependence} is exactly removed. Moreover, due to the independence on the renormalization scheme and scale, the resulting conformal series with an overall single value of $\alpha_s(Q_*)$ provides not only precise pQCD predictions for the known fixed order, but also a reliable basis for estimating the contributions from the unknown higher-order terms.

Within the framework of the pQCD, the perturbative approximant for physical observable $\varrho$ can be written in the following form:
\begin{eqnarray}
\varrho_{n} = \sum^{n}_{i=1} r_{i}(\mu^2_r/Q^2) \alpha_s^{p+i-1}(\mu_r), \label{eq:rho}
\end{eqnarray}
where $Q$ represents the kinematic scale and the index $p(p\ge1)$ indicates the $\alpha_s$-order of the leading-order (LO) contribution. For the perturbative series (\ref{eq:rho}), its perturbative coefficients $r_i$ can be divided into the conformal parts ($r_{i,0}$) and non-conformal parts (proportional to $\beta_i$), i.e. $r_i=r_{i,0}+\mathcal{O}(\{\beta_i\})$. The $\{\beta_i\}$-pattern at different orders exist a special degeneracy~\cite{Brodsky:2013vpa, Mojaza:2012mf,Bi:2015wea}, i.e.
\begin{eqnarray}
r_1 &=& r_{1,0}, \nonumber\\
r_2 &=& r_{2,0} + p \beta_0 r_{2,1}, \nonumber\\
r_3 &=& r_{3,0} + p \beta_1 r_{2,1} + (p+1){\beta _0}r_{3,1} + \frac{p(p+1)}{2} \beta_0^2 r_{3,2}, \nonumber\\
r_4 &=& r_{4,0} + p{\beta_2}{r_{2,1}} + (p+1){\beta_1}{r_{3,1}} + \frac{p(3+2p)}{2}{\beta_1}{\beta_0}{r_{3,2}} \nonumber\\
&& + (p+2){\beta_0}{r_{4,1}}+ \frac{(p+1)(p+2)}{2}\beta_0^2{r_{4,2}} \nonumber\\
&& + \frac{p(p+1)(p+2)}{3!}\beta_0^3{r_{4,3}}, \nonumber\\
&& \hspace{-4mm} \cdots \nonumber
\end{eqnarray}
The coefficients $r_{i,j}$ are general functions of the renormalization scale $\mu_r$, which can be redefined as
\begin{eqnarray}
r_{i,j}=\sum^j_{k=0}C^k_j{\hat r}_{i-k,j-k}{\rm ln}^k(\mu_r^2/Q^2),~\label{rijrelation}
\end{eqnarray}
where the reduced coefficients ${\hat r}_{i,j}=r_{i,j}|_{\mu_r=Q}$, and the combination coefficients $C^k_j=j!/(k!(j-k)!)$.

Following the standard PMCs procedures~\cite{Shen:2017pdu}, the overall effective scale can be determined by requiring all the nonconformal $\{\beta_i\}$-terms to vanish; the pQCD approximant (\ref{eq:rho}) then changes to the following conformal series,
\begin{eqnarray}
\varrho_n|_{\rm PMCs}=\sum_{i=1}^n \hat{r}_{i,0}\alpha_s^{p+i-1}(Q_{*}),
\label{pmcs}
\end{eqnarray}
where the PMC scale $Q_{*}$ can be fixed up to N$^2$LL-accuracy for $n=4$, i.e. $\ln Q^2_* / Q^2$ can be expanded as a power series over $\alpha_s(Q)$,
\begin{eqnarray}
\ln\frac{Q^2_*}{Q^2}=T_0+T_1 \alpha_s(Q)+T_2 \alpha_s^2(Q)+ \mathcal{O}(\alpha_s^3),
\label{qstar}
\end{eqnarray}
where the coefficients $T_i~(i=0, 1, 2)$ are all functions of the reduced coefficients ${\hat r}_{i,j}$, whose expressions can be found in Ref.~\cite{Shen:2017pdu}. Equation (\ref{qstar}) shows that the PMC scale $Q_*$ is also represented as power series in $\alpha_s$, which resums all the known $\{\beta_i\}$-terms, and is explicitly independent of $\mu_r$ at any fixed order. It represents the physical momentum flow of the process and determines an overall effective value of $\alpha_s$. Together with the $\mu_r$-independent conformal coefficients, the resulting pQCD series is exactly scheme-and-scale independent~\cite{Wu:2018cmb}, thus providing a reliable basis for estimating the contributions of the unknown terms.

\section{Numerical results}
\label{sec:numer}

In this section, we apply the PMCs approach to scale setting in combination with the Bayesian method for estimating uncertainties from the uncalculated higher order terms, for two physical observables $R_{e^+e^-}$ and $R_{\tau}$, all of which are now known up to four loops in pQCD. We will show how the magnitude of the ``unknown'' terms predicted by the Bayesian-based approach varies as more-and-more loop terms are determined.

The ratio $R_{e^+e^-}$ for $e^+e^-$ annihilation is defined as
\begin{eqnarray}
R_{e^+ e^-}(Q) &=& \frac{\sigma\left(e^+e^-\to {\rm hadrons} \right)}{\sigma\left(e^+e^-\to \mu^+ \mu^-\right)}\nonumber\\
&=& 3\sum_q e_q^2\left[1+R(Q)\right], \label{ree}
\end{eqnarray}
where $Q=\sqrt{s}$ is the $e^+e^-$ center-pf-mass collision energy at which the ratio is measured. The pQCD approximant of $R(Q)$, denoted by $R_n(Q)$, reads, $R_n(Q)= \sum_{i=1}^{n} r_i(\mu_r/Q) \alpha_s^{i}(\mu_r)$. The pQCD coefficients at $\mu_r=Q$ have been calculated in the $\overline{\rm MS}$-scheme in Refs.~\cite{Baikov:2008jh, Baikov:2010je, Baikov:2012zm, Baikov:2012zn}. The coefficients at any other scales can then be obtained via RGE evolution. For illustration, we shall take $Q\equiv 31.6 \;{\rm GeV}$~\cite{Marshall:1988ri} throughout this paper to illustrate the numerical predictions.

The ratio $R_{\tau}$ for hadronic $\tau$ decays is defined as
\begin{eqnarray}
R_{\tau}(M_{\tau}) &=&\frac{\sigma(\tau\rightarrow\nu_{\tau}+\rm{hadrons)}}{\sigma(\tau\rightarrow\nu_{\tau}+\bar{\nu}_e+e^-)}\nonumber\\
&=&3\sum\left|V_{ff'}\right|^2\left(1+\tilde{R}(M_{\tau})\right),
\end{eqnarray}
where $V_{ff'}$ are Cabbibo-Kobayashi-Maskawa matrix elements, $\sum\left|V_{ff'}\right|^2 =\left|V_{ud}\right|^2+\left|V_{us}\right|^2\approx 1$ and $M_{\tau}= 1.77686$ GeV \cite{Workman:2022ynf}. The pQCD approximant of $\tilde{R}(M_{\tau})$, denoted by $\tilde{R}_{n}(M_{\tau})$, reads, $\tilde{R}_{n}(M_{\tau})= \sum_{i=1}^{n}r_i(\mu_r/M_{\tau})\alpha_s^{i}(\mu_r)$; the coefficients can be obtained using the known relation of $R_{\tau}(M_{\tau})$ to $R_{e^+ e^-}(Q)$~\cite{Lam:1977cu}.

In order to do the numerical evaluation, the RunDec program \cite{Chetyrkin:2000yt, Herren:2017osy} is adopted to calculate the value of $\alpha_s$. For self-consistency, the four-loop $\alpha_s$-running behavior will be used. The world average $\alpha_s(M_z)=0.1179\pm 0.0009$~\cite{Workman:2022ynf} is adopted as a reference.

\subsection{Single-scale PMCs predictions}

After applying the PMCs approach, the overall renormalization scale for each process can be determined. If the pQCD approximants are known up to two-loop, three-loop, and four-loop level, respectively, the corresponding overall scales are
\begin{eqnarray}
Q_{*}|_{e^+e^-} &=& \{35.36, 39.67, 40.28\} \; {\rm GeV}, \\
Q_{*}|_{\tau} &=& \{0.90, 1.01, 1.05 \} \; {\rm GeV}.
\end{eqnarray}
The PMC scales $Q_*$ are independent of the initial choice of the renormalization scale $\mu_r$. In the case of the leading-order ratios with $n=1$, one has no information to set the effective scale, and thus for definiteness, we will set it to be $Q$, or $M_\tau$, respectively, which gives $R_1=0.04428$, and $\tilde{R}_1=0.0891$.

\begin{table}[htb]
\caption{The known coefficients for $R_n(Q)$. The coefficients of conventional scale setting, $r_i(\mu_r)$, are for $\mu_r=Q$. The conformal coefficients, $r_{i,0}$, are scale-independent}
\begin{tabular}{ c c c c c }
\hline
 & ~~$i=1$~~ & ~~$i=2$~~ & ~~$i=3$~~ & ~~$i=4$~~ \\ \hline
$r_i(\mu_r=Q)$ & $0.3183$ & $0.1428$ & $-0.4130$ & $0.8257$ \\
$r_{i,0}$ & $0.3183$ & $0.1865$ & $-0.0324$ & $-0.1128$ \\
\hline
\end{tabular}
\label{tab:ReeCoeff}
\end{table}

\begin{table}[htb]
\caption{The known coefficients for $\tilde{R}_n(M_\tau)$. The coefficients of conventional scale setting, $r_i(\mu_r)$, are for $\mu_r=M_\tau$. The conformal coefficients, $r_{i,0}$, are scale-independent}
\begin{tabular}{ c c c c c }
\hline
& ~~$i=1$~~ & ~~$i=2$~~ & ~~$i=3$~~ & ~~$i=4$~~ \\ \hline
$r_i(\mu_r=M_\tau)$ & $0.3183$ & $0.5271$ & $0.8503$ & $1.3046$ \\
$r_{i,0}$ & $0.3183$ & $0.2174$ & $0.1108$ & $0.0698$ \\
\hline
\end{tabular}
\label{tab:RtauCoeff}
\end{table}

We present the first four conformal coefficients $r_{i,0}$ ($i=1,2,3,4$) in Tables \ref{tab:ReeCoeff} and \ref{tab:RtauCoeff}, in which the conventional coefficients $r_i$ ($i=1,2,3,4$) at a specified scale are also presented in comparison. Because the coefficients $r_i (i\geq 2)$ of the conventional pQCD series are scale-dependent at every orders, the Bayesian-based approach can only be applied after one specifies the choices for the renormalization scale, thus introducing extra uncertainties for the Bayesian-based approach. On the other hand, the PMCs series is a scale-independent conformal series in powers of the effective coupling $\alpha_s(Q_*)$; the PMCs thus provides a reliable basis for obtaining constraints on the predictions for the unknown higher-order contributions.

\subsection{Estimation of UHOs using the Bayesian-based approach}

In this subsection, we give estimates for the UHOs of the pQCD series $R_n(Q=31.6\;{\rm GeV})$ and $\tilde{R}_n(M_\tau)$. More explicitly, we will predict the magnitude of the unknown coefficient $c_{i+1}$ from the known ones $\{c_{1},\cdots,c_{i}\}$ by using the Bayesian-based approach.

\begin{table}[htb]
\caption{The predicted smallest $95.5\%$ credible intervals (CI) for the scale-dependent conventional coefficients $r_i(\mu_r)$ ($i=3,4,5$) at the scale $\mu_r=Q$ and the scale-invariant coefficients $r_{i,0} (i=3,4,5)$ of $R_n(Q=31.6\;{\rm GeV})$ via the Bayesian-based approach. The exact values (``EC'') are presented as comparisons}
\begin{tabular}[b]{cccc}
\hline
 & ~~$r_3(\mu_r=Q)$~~ & ~~$r_4(\mu_r=Q)$~~ & ~~$r_5(\mu_r=Q)$~~ \\
\hline
CI & $[-0.8663,0.8663]$ & $[-0.7314,0.7314]$ & $[-1.1989,1.1989]$ \\
EC & $-0.4130$ & $0.8257$ & - \\
\hline
 & ~~$r_{3,0}$~~ & ~~$r_{4,0}$~~ & ~~$r_{5,0}$~~ \\
\hline
CI & $[-0.8663,0.8663]$ & $[-0.5638,0.5638]$ & $[-0.4622,0.4622]$ \\
EC & $-0.0324$ & $-0.1128$ & - \\
\hline
\end{tabular}
\label{tab:Reeri0}
\end{table}

\begin{table}[htb]
\caption{The predicted smallest $95.5\%$ credible intervals (CI) for the scale-dependent conventional coefficients $r_i(\mu_r)$ ($i=3,4,5$) at the scale $\mu_r=M_\tau$ and the scale-invariant coefficients $r_{i,0} (i=3,4,5)$ of $\tilde{R}_n(M_{\tau})$ via the Bayesian-based approach. The exact values (``EC'') are presented for comparison}
\begin{tabular}[b]{cccc}
\hline
 & ~~$r_3(\mu_r=M_\tau)$~~ & ~~$r_4(\mu_r=M_\tau)$~~ & ~~$r_5(\mu_r=M_\tau)$~~ \\
 \hline
CI & $[-1.4346,1.4346]$ & $[-1.5060,1.5060]$ & $[-1.8942,1.8942]$ \\
EC & $0.8503$ & $1.3046$ & - \\ \hline
 & ~~$r_{3,0}$~~ & ~~$r_{4,0}$~~ & ~~$r_{5,0}$~~ \\ \hline
CI & $[-0.8663,0.8663]$ & $[-0.5638,0.5638]$ & $[-0.4622,0.4622]$ \\
EC & $0.1108$ & $0.0698$ & - \\ \hline
\end{tabular}
\label{tab:Rtauri0}
\end{table}

\begin{table}[htb]
\caption{The predicted smallest $95.5\%$ credible intervals (CI) for the scale-dependent coefficients $r_3(\mu_r)$ and $r_4(\mu_r)$ of $R_n(Q=31.6\;{\rm GeV})$ by using the Bayesian-based approach at three scales $\mu_r=Q$, $Q/2$ and $2Q$, respectively. The exact values (``EC'') are present for comparison}
\begin{tabular}[b]{cccccc}
\hline
& & ~~$r_3(\mu_r)$~~ & ~~$r_4(\mu_r)$~~ \\ \hline
$\mu_r=Q$ & CI & $[-0.8663,0.8663]$ & $[-0.7314,0.7314]$ \\
& EC & $-0.4130$ & $-0.8257$ \\ \hline
$\mu_r=2Q$ & CI & $[-1.1213,1.1213]$ & $[-0.7297,0.7297]$ \\
& EC & $0.1643$ & $-1.0089$ \\ \hline
$\mu_r=Q/2$ & CI & $[-0.8663,0.8663]$ & $[-0.9473,0.9473]$ \\
& EC & $-0.5348$ & $0.4272$ \\ \hline
\end{tabular}
\label{tab:ri}
\end{table}

\begin{figure*}[htb]
\includegraphics[width=0.45\textwidth]{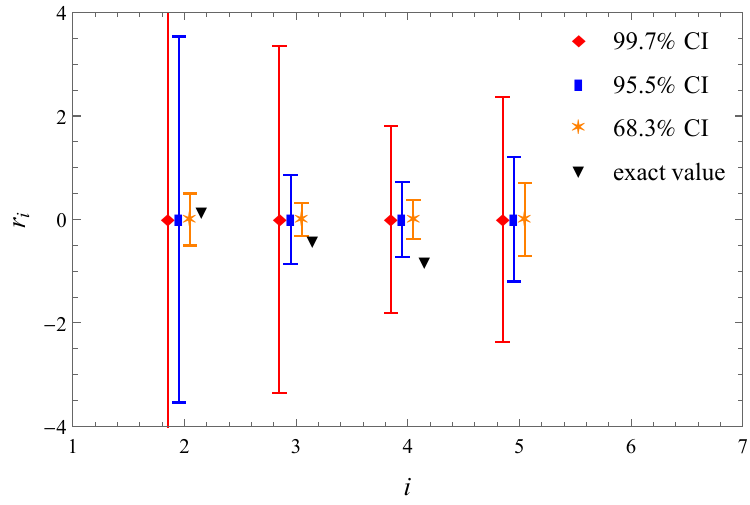}~~~
\includegraphics[width=0.45\textwidth]{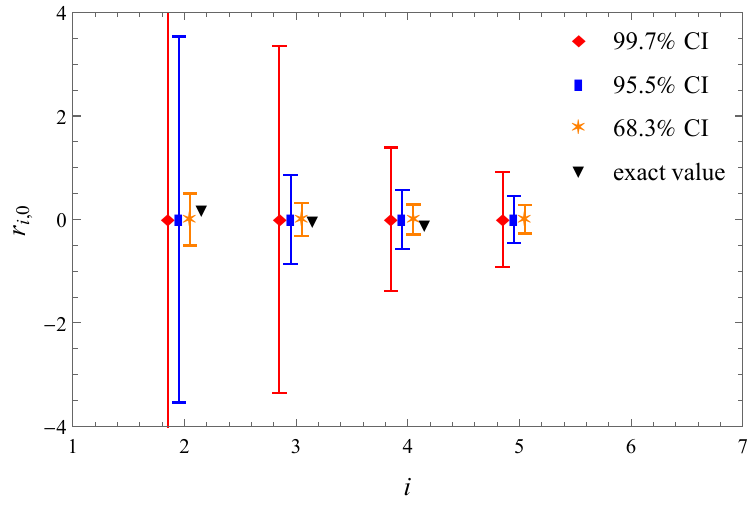}
\caption{The predicted credible intervals (CI) with three typical DoBs for the scale-dependent coefficients $r_i(\mu_r)$ at the scale $\mu_r=Q$ and the scale-invariant $r_{i,0}$ of $R_n(Q=31.6{\rm GeV})$ under the Bayesian-based approach, respectively. The red diamonds, the blue rectangles, the golden yellow stars and the black inverted triangles together with their error bars, are for $99.7\%$ CI, $95.5\%$ CI, $68.3\%$ CI, and the exact values of the coefficients at different orders, respectively}
\label{Fig:ReeCoefficients}
\end{figure*}

\begin{figure*}[htb]
\includegraphics[width=0.45\textwidth]{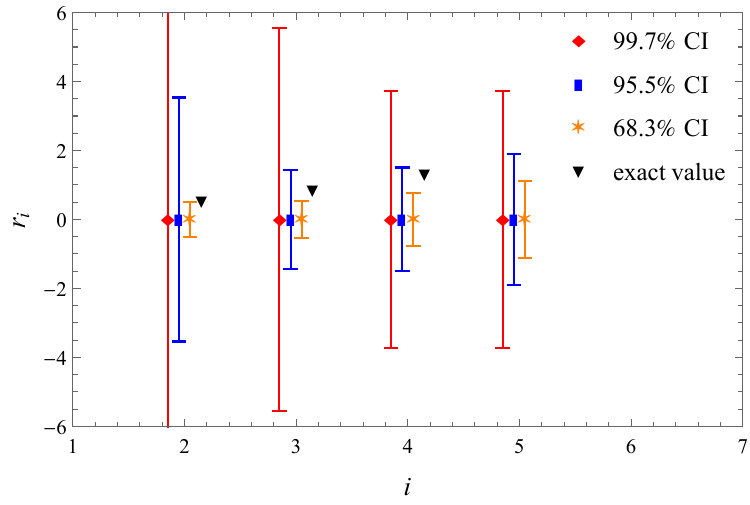}~~~
\includegraphics[width=0.45\textwidth]{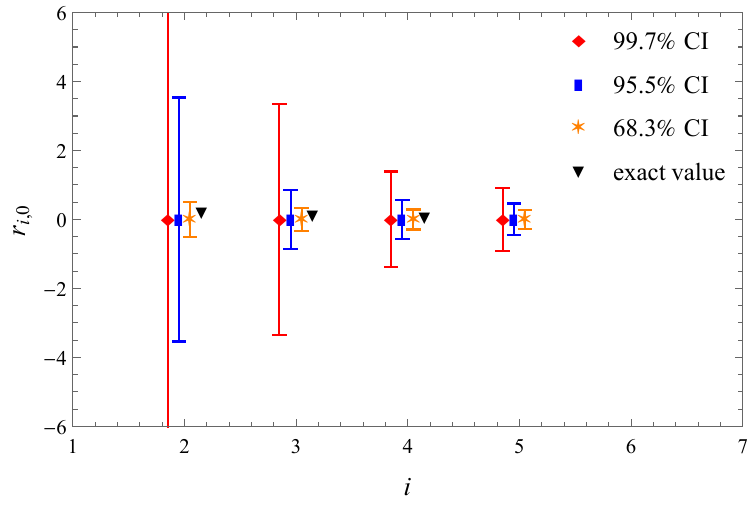}
\caption{The predicted credible intervals (CI) with three typical DoBs for the scale-dependent coefficients $r_i(\mu_r)$ at the scale $\mu_r=M_\tau$ and the scale-invariant $r_{i,0}$ of $\tilde{R}_n(M_\tau)$ under the Bayesian-based approach, respectively. The red diamonds, the blue rectangles, the golden yellow stars and the black inverted triangles together with their error bars, are for $99.7\%$ CI, $95.5\%$ CI, $68.3\%$ CI, and the exact values of the coefficients at different orders, respectively}
\label{Fig:RtauCoefficients}
\end{figure*}

\begin{figure*}[htb]
\includegraphics[width=0.45\textwidth]{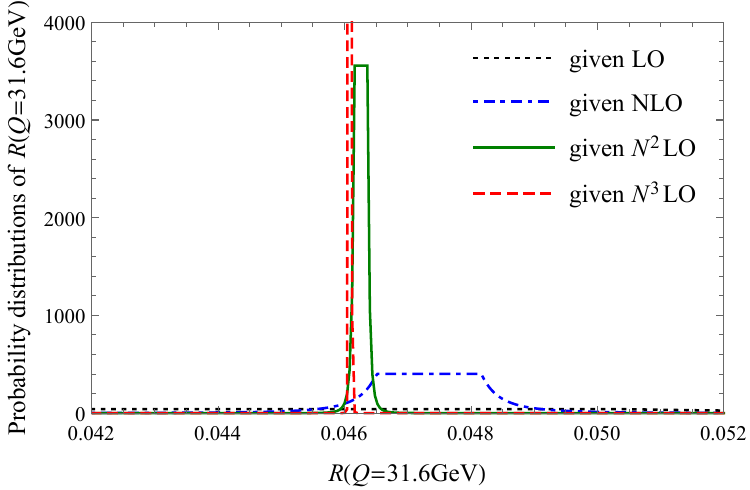}~
\includegraphics[width=0.45\textwidth]{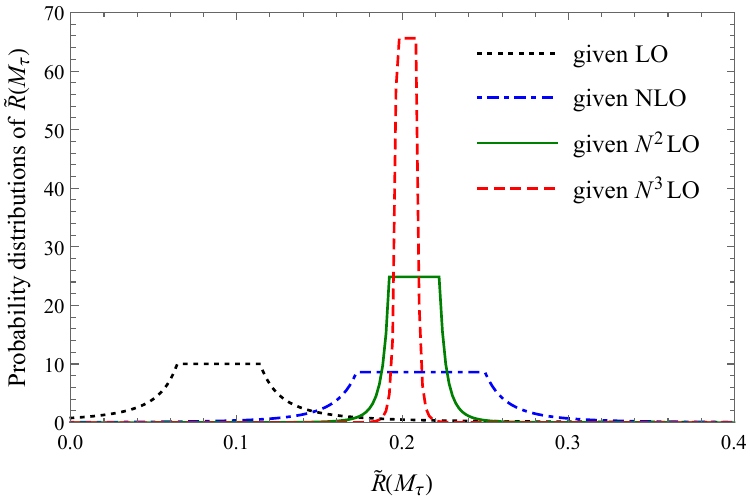}~
\caption{The probability density distributions of $R(Q=31.6\;{\rm GeV})$ and $\tilde{R}(M_\tau)$ with different states of knowledge predicted by PMCs and the Bayesian-based approach, respectively. The black dotted, the blue dash-dotted, the green solid and the red dashed lines are results for the given LO, NLO, N$^2$LO and N$^3$LO series, respectively}
\label{fig:Reedistribution}
\end{figure*}

\begin{figure*}[htb]
\includegraphics[width=0.45\textwidth]{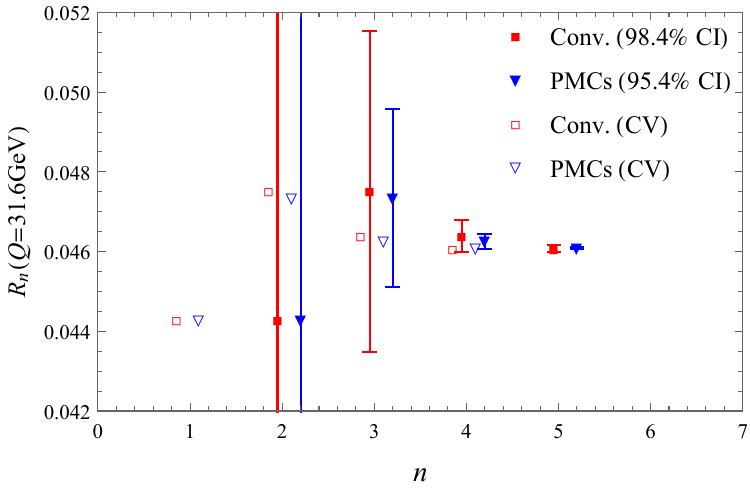}~
\includegraphics[width=0.45\textwidth]{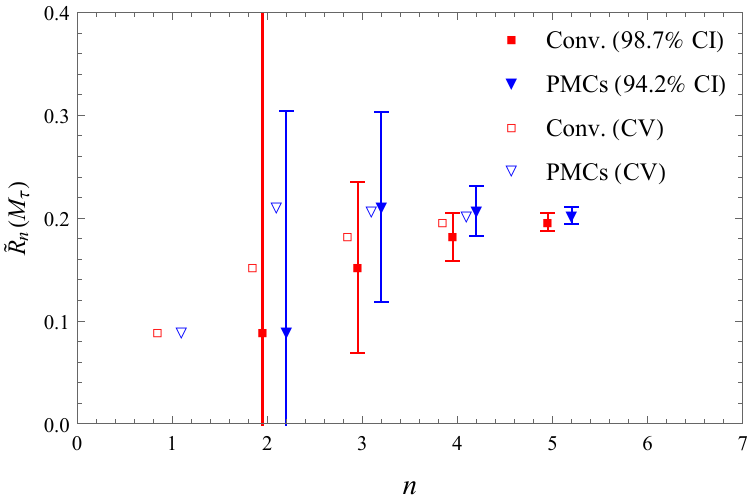}~
\caption{Comparison of the calculated central values (``CV'') of the pQCD approximants $R_n(Q=31.6\;{\rm GeV})$ and $\tilde{R}_n(M_\tau)$ for $n=(1,2,3,4)$ with the predicted $p_s\%$ credible intervals (CI) of those approximants for $n=(2,3,4,5)$. The blue hollow triangles and red hollow quadrates represent the calculated central values of the fixed-order pQCD predictions using PMCs and conventional (Conv.) scale-setting, respectively. The blue solid triangles and red solid quadrates with error bars represent the predicted $p_s\%$ CIs using the Bayesian-based approach based on the PMCs conformal series ($p_s=95.4$ for $R_n(Q=31.6\;{\rm GeV})$, $p_s=94.2$ for $\tilde{R}_n(M_\tau)$) and the conventional (Conv.) scale-dependent series ($p_s=98.4$ for $R_n(Q=31.6\;{\rm GeV})$, $p_s=98.7$ for  $\tilde{R}_n(M_\tau)$), respectively}
\label{fig:ReeRn}
\end{figure*}

\begin{table*}[htb]
\caption{Comparison of the calculated central values (``CV'') with the predicted $p_s\%$ credible intervals (``CI'') of the pQCD approximants $R_n(Q=31.6{\rm GeV})$ and $\tilde{R}_{n}(M_\tau)$ based on the PMC conformal series and the conventional (conv.) scale-dependent series up to $n_{\rm th}$-order level, respectively. The DoB of the CI ($p_s\%$) is given in the last column. The results for the PMC conformal series are scale-independent. The results for the conventional scale-dependent series are calculated at $\mu_r=Q$ and $M_\tau$ for $R_n(Q=31.6{\rm GeV})$ and $\tilde{R}_{n}(M_\tau)$, respectively}
\centering
\begin{tabular}{cccccccc}
\hline
 &~${\rm CV}$, $n=2$~&~${\rm CI}$, $n=3$~&~${\rm CV}$, $n=3$~&~${\rm CI}$, $n=4$~&~ ${\rm CV}$, $n=4$~&~${\rm CI}$, $n=5$~&~$p_s\%$~ \\
 \hline
$R_n(Q=31.6{\rm GeV})|_{\rm PMCs}$ & $0.04733$ & $[0.04510,0.04956]$ & $0.04626$ & $[0.04607,0.04645]$ & $0.04609$ & $[0.04607,0.04611]$ & $95.4\%$ \\
$\tilde{R}_{n}(M_{\tau})|_{\rm PMCs}$ & $0.2110$ & $[0.1183,0.3037]$ & $0.2071$ & $[0.1825,0.2317]$ & $0.2023$ & $[0.1940,0.2106]$ & $94.2\%$ \\
$R_n(Q=31.6{\rm GeV})|_{\rm conv.}$ & $0.04751$ & $[0.04349,0.05153]$ & $0.04639$ & $[0.04599,0.04679]$ & $0.04607$ & $[0.04598,0.04616]$ & $98.4\%$ \\
$\tilde{R}_{n}(M_{\tau})|_{\rm conv.}$ & $0.1523$ & $[0.0692,0.2354]$ & $0.1820$ & $[0.1586,0.2054]$ & $0.1965$ & $[0.1879,0.2051]$ & $98.7\%$ \\
\hline
\end{tabular}
\label{tab:Total-observable}
\end{table*}

First, we present the predicted smallest $95.5\%$ CIs and the exact values~\footnote{The ``exact" value means that it is obtained by directly using the known perturbative series.} (``EC'') of the scale-invariant conformal coefficients $c_i=r_{i,0}$ $(i=3,4,5)$ of the PMCs series of $R_n(Q=31.6\;{\rm GeV})$ and $\tilde{R}_n(M_{\tau})$ in Tables \ref{tab:Reeri0} and \ref{tab:Rtauri0}, respectively. For comparison, the similarly predicted scale-dependent conventional coefficients $c_i=r_i(\mu_r)$ $(i=3,4,5)$ of the conventional series of $R_n(Q=31.6\;{\rm GeV})$ and $\tilde{R}_n(M_{\tau})$ at the specific scales $\mu_r=Q$ and $M_\tau$ are also presented. It is noted that the exact values of $r_{3,0}$ and $r_{4,0}$ lie within the predicted $95.5\%$ CIs. In contrast, for the conventional coefficients, most of the exact values of $r_{3}$ and $r_{4}$ are lying within the predicted $95.5\%$ CIs; However, there are one exception for $r_4$, i.e. for $R_n(Q=31.6\;{\rm GeV})$, the exact values of $r_4$ are outside the region of the $95.5\%$ CIs. These exceptions may be removed by taking a different choice of renormalization scale; e.g., for the case of $R_n(Q=31.6\;{\rm GeV})$, as shown by Table \ref{tab:ri}, the exact value of $r_{4}$ will lie within the predicted $95.5\%$ CI if setting $\mu_r=Q/2$. Table \ref{tab:ri} also confirms that the CIs predicted from the conventional series are also scale dependent. Thus, in comparison with the renormalon-divergent and scale-dependent conventional series, it is essential to use the renormalon-free and scale-invariant PMCs series in order to estimate the unknown higher-order coefficients. More explicitly, we present more predicted CIs with three typical DoBs in Figs.~\ref{Fig:ReeCoefficients} and \ref{Fig:RtauCoefficients}, respectively.

Second, we present the probability density distributions for the two observables $R(Q=31.6\;{\rm GeV})$ and $\tilde{R}(M_\tau)$ with different states of knowledge predicted by PMCs and the Bayesian-based approach in Fig.~\ref{fig:Reedistribution}. The four lines in each figure correspond to different degrees of knowledge: given LO (dotted), given NLO (dotdashed), given N$^2$LO (solid) and given N$^3$LO (dashed). These figures illustrate the characteristics of the posterior distribution: a symmetric plateau with two suppressed tails. The posterior distribution given by the Bayesian-based approach depends on the prior distribution, and as more and more loop terms become known, the probability is updated with less and less dependence on the prior; i.e., the probability density becomes increasingly concentrated (the plateau becomes narrower and narrower and the tail becomes shorter and shorter) as more and more loop terms for the distribution are determined.

Third, we present the $p_s\%$ CIs of $R_n(Q=31.6\;{\rm GeV})$ with $n=(2,3,4,5)$ predicted from the one-order lower $R_{n-1}(Q=31.6\;{\rm GeV})$ based on the Bayesian-based approach in Fig. \ref{fig:ReeRn}, where $p_s\%=95.4\%$ for the scale-independent PMCs series, and $p_s\%=98.4\%$ for the scale-dependent conventional series. The calculated values (``CV'') of the pQCD approximants $R_n(Q=31.6\;{\rm GeV})$ with $n=(1,2,3,4)$ are also presented as a comparison. The triangles and the quadrates are for the PMCs series and the conventional (conv.) scale-dependent series, respectively. Analogous results for $\tilde{R}_{n}(M_\tau)$ are also given in Fig. \ref{fig:ReeRn}. Both the center values and the error bars (or CIs) are scale-independent for the PMCs series. The results for conventional series of $R_n$ and $\tilde{R}_{n}$ are for $\mu_r=Q$ and $M_\tau$, respectively. Fig. \ref{fig:ReeRn} shows that the error bars (or CIs) predicted by using the Bayesian-based approach quickly approach their steady points for both the PMCs and conventional series. As expected, the error bars provide consistent and high DoB estimates for the UHOs for both the PMCs and conventional series; e.g., the error bars of $R_{n+1}(Q)$ ($n=2,3,4$) predicted from $R_n(Q)$ are well within the error bars of the one-order lower $R_n(Q)$ predicted from $R_{n-1}(Q)$; The conclusions for $\tilde{R}_{n}(M_\tau)$ are similar. Detailed numerical results are presented in Table~\ref{tab:Total-observable}, where the 2nd, 4th, and 6th columns show the calculated central values (``CV'') of the fixed-order pQCD approximants $R_n(Q=31.6{\rm GeV})$ and $\tilde{R}_{n}(M_\tau)$ for $n=2,3,4$ respectively, and the 3rd, 5th, and 7th columns show the predicted $p_s\%$ credible intervals (``CI'') of those approximants for $n=3,4,5$ respectively. The predicted CIs for $R_2(Q)$ and $\tilde{R}_2(M_\tau)$ are sufficiently conservative and thus are not presented in the table. The DoB ($p_s\%$) is given in the last column. For the present prior distributions, $p_\sigma\%=65.3\%$ for $l=1$ and $k=4$. Thus the DoB $p_s\%$, given in Table~\ref{tab:Total-observable}, is also the critical DoB, i.e. $p_s=p_c$.

Our final predictions for the five-loop predictions of $R_5(Q)$ and $\tilde{R}_5(M_\tau)$ based on the PMCs and the Bayesian-based approach read,
\begin{eqnarray}\label{eq:PMCs prediction}
R_5(Q=31.6{\rm GeV}) &=& 0.04609\pm0.00042\pm0.00002, \\
\tilde{R}_5(M_\tau) &=& 0.2032^{+0.0092}_{-0.0086}\pm0.0083,
\end{eqnarray}
where the first error is for $\Delta\alpha_s(M_Z)=\pm 0.0009$ and the second error represents high DoBs $p_s\%$ which are consistent with the estimates for the UHOs. Note that the very small uncertainty $\pm 0.00002 $ for $R_5(Q=31.6{\rm GeV})$ is determined by the $95.4\%$ CI according to the Bayesian-based approach, $[-r_{5,0}^{(95.4)}\alpha_s^5(Q_*),r_{5,0}^{(95.4)}\alpha_s^5(Q_*)]$, where $\alpha_s^5(Q_*)\simeq 0.00004$ and the predicted $r_{5,0}^{(95.4)}=0.4596$ are all small. Our prediction for hadronic $\tau$ decays, $\tilde{R}_5(M_\tau)$, can be compared with those given in Refs.\cite{Beneke:2008ad, Boito:2016pwf, Boito:2018rwt, Caprini:2018agy}.

\section{Summary}
\label{sec:summary}

The PMC provides a rigorous first-principle method to eliminate conventional renormalization scheme and scale ambiguities for high-momentum transfer processes in pQCD up to any fixed order. Its predictions have a solid theoretical foundation, satisfying renormalization group invariance and all other self-consistency conditions derived from the renormalization group. The PMCs is a single-scale-setting approach, which determines a single overall effective/correct $\alpha_s(Q_*)$ by using all of the RG-involved nonconformal $\{\beta_i\}$-terms. The resulting PMCs series is a renormalon-free and scale-invariant conformal series; it thus achieves precise fixed-order pQCD predictions and provides a reliable basis for predicting unknown higher-order contributions.

The Bayesian analysis provides a compelling approach for estimating the UHOs from the known fixed-order series by adopting a probabilistic interpretation. The conditional probability of the unknown perturbative coefficient is first given by a subjective prior distribution, which is then updated iteratively according to the Bayes' theorem as more and more information is included. The posterior distribution given by the Bayesian-based approach depends on the subjective prior distribution (or the assumptions), and as more-and-more information updates the probability, less-and-less dependence on the prior distribution (or the assumptions) can be achieved, as confirmed in Fig. \ref{fig:Reedistribution}.

We have defined an objective measure which characterizes the uncertainty due to the uncalculated higher order (UHO) contributions of a perturbative QCD series using the Bayesian analysis. This uncertainty is given as a credible interval (CI) with a degree of belief (DoB, also called Bayesian probability). The numerical value for the uncertainty,  the critical DoB,  is given as a percentage $p_c \%$.  When $p_c\% =95\%$, it means that there is a $95\%$ probability that the exact answer is within this range. The CI with DoB $p_s \%$ in Fig. \ref{fig:ReeRn} and Table~\ref{tab:Total-observable} takes into account the uncertainties in the values of the input physics parameters, such as the value of $\alpha_s$, which will become very small at high order due to the $\alpha_s^n$-power suppression. Detailed numerical results are presented in Table~\ref{tab:Total-observable}, where the 2nd, 4th, and 6th columns show the calculated central values of the fixed-order pQCD approximants $R_n(Q=31.6{\rm GeV})$ and $\tilde{R}_{n}(M_\tau)$ for $n=2,3,4$, respectively, and the 3rd, 5th, 7th columns show the predicted $p_s\%$ credible interval of those approximants for $n=3,4,5$, respectively. The 8th column shows the DoB ($p_s\%$) of the credible interval presented in the 3rd, 5th, 7th columns. The calculated $p_s$ value, $p_s={\rm max}\{p_c,p_\sigma\}$, equals $p_c$ since the DoB of the $1\sigma$-interval $p_\sigma\%$ equals $65.3\%$ for the present prior distributions.

In contrast, each term in a conventional perturbative series is highly scale-dependent, thus the Bayesian-based approach can only be applied after one assumes choices for the perturbative scale. What's more, the $n !$ renormalon series leads to divergent behavior especially at high order~\footnote{Such renormalon divergence also makes the hidden parameter $\bar{c}$ to be much larger than the PMC one, thus if choosing the same degree-of-belief, the PMC credible interval shall be much smaller. }; e.g., the exact value of the conventional coefficient $r_4$ is even outside the $95.5\%$ CI predicted from $\{r_1,r_2,r_3\}$ for $R(Q)$, which can be found in Table \ref{tab:Reeri0}. Thus, it is critical to use the more convergent and scale-independent PMC conformal series as the basis for estimating the unknown higher-order coefficients.

As we have shown, by using the PMCs approach in combination with the Bayesian analysis, one can obtain highly precise fixed-order pQCD predictions and achieve consistent estimates with high DoB for the unknown higher-order contributions. In the present paper, we have illustrated this procedure for two important hadronic observables, $R_{e^+e^-}$ and $R_{\tau}$, which have been calculated up to four-loops in pQCD. The elimination of the uncertainty in setting the renormalization scale for fixed-order pQCD predictions using the PMCs, together with the reliable estimate for the uncalculated higher-order contributions obtained using the Bayesian analysis, greatly increases the precision of collider tests of the Standard Model and thus the sensitivity to new phenomena.

\hspace{1cm}

\noindent{\bf Acknowledgments}: This work was supported in part by the Natural Science Foundation of China under Grant No.11905056, No.12147102 and No.12175025, by the graduate research and innovation foundation of Chongqing, china (No.CYB21045 and No.ydstd1912), and by the Department of Energy (DOE), Contract DECAC02C76SF00515. SLAC-PUB-17690.

\appendix

\section{Theorems and laws for probability distributions}

An abstract definition of probability can be given by considering a set $S$, called the sample space, and possible subsets $A,B,\cdots$, the interpretation of which is left open. The \emph{probability} $P$ is a real-valued function defined by the following axioms due to Kolmogorov \cite{Kolmogorov:1933, Kolmogorov:1956}:
\begin{enumerate}
 \item For every subset $A$ in $S$, $P(A)\geq 0$;
 \item For disjoint subsets (i.e., $A\cap B=\emptyset$), \\ $P(A\cup B)=P(A)+P(B)$;
 \item $P(S)=1$.
\end{enumerate}
In addition, the \emph{conditional probability} $P(A|B)$ (read as $P$ of $A$ given $B$) is defined as
\begin{eqnarray}\label{eq:cpdf}
P(A|B)=\frac{P(A\cap B)}{P(B)}.
\end{eqnarray}
From this definition and using the fact that $A\cap B$ and $B\cap A$ are the same, one obtains \emph{Bayes' theorem},
\begin{eqnarray}\label{eq:bayes0}
P(A|B)=\frac{P(B|A)P(A)}{P(B)}.
\end{eqnarray}
From the three axioms of probability and the definition of conditional probability, one obtains the \emph{law of total probability},
\begin{eqnarray}\label{eq:law}
P(B)=\sum_i P(B|A_i)P(A_i),
\end{eqnarray}
for any subset $B$ and for disjoint $A_i$ with $\cup_iA_i=S$. This can be combined with Bayes' theorem (\ref{eq:bayes0}) to give
\begin{eqnarray}\label{eq:bayes1}
P(A|B)=\frac{P(B|A)P(A)}{\sum_i P(B|A_i)P(A_i)},
\end{eqnarray}
where the subset $A$ could, for example, be one of the $A_i$.

\section{The Bayesian analysis}

In Bayesian statistics, the subjective interpretation of probability is used to quantify one's \emph{degree of belief} in a \emph{hypothesis}. The hypothesis is often characterized by one or more parameters. This allows one to define a \emph{probability density function} (p.d.f.) for a parameter, which reflects one's knowledge about where its true value lies.

Consider an experiment whose outcome is characterized by a vector of data $\pmb{x}$. A hypothesis $H$ is a statement about the probability for the data, often written $P(\pmb{x}|H)$. This could, for example, completely define the p.d.f. for the data (a simple hypothesis), or it could specify only the functional form of the p.d.f., with the values of one or more parameters not determined (a composite hypothesis). If the probability $P(\pmb{x}|H)$ for data $\pmb{x}$ is regarded as a function of the hypothesis $H$, then it is called the \emph{likelihood} of $H$, usually written $L(H)$. Consider the hypothesis $H$ is characterized by one or more continuous parameters $\pmb{\theta}$, in which case $L(\pmb{\theta})=P(\pmb{x}|\pmb{\theta})$ is called the \emph{likelihood function}. Note that the likelihood function itself is not a p.d.f. for $\pmb{\theta}$.

In the Bayesian analysis, inference is based on the posterior p.d.f. $p(\pmb{\theta}|\pmb{x})$, whose integral over any given region gives the degree of belief for $\pmb{\theta}$ to take on values in that region, given the data $\pmb{x}$. This is obtained from Bayes' theorem (\ref{eq:bayes1}), which can be written
\begin{eqnarray}\label{eq:bayes2}
p(\pmb{\theta}|\pmb{x})=\frac{P(\pmb{x}|\pmb{\theta})\pi(\pmb{\theta})}{\int P(\pmb{x}|\pmb{\theta}')\pi(\pmb{\theta}') d \pmb{\theta}'},
\end{eqnarray}
where $P(\pmb{x}|\pmb{\theta})$ is the likelihood function for $\pmb{\theta}$; i.e., the joint p.d.f. for the data viewed as a function of $\pmb{\theta}$, evaluated with data actually obtained in the experiment. The function $\pi(\pmb{\theta})$ is the prior p.d.f. for $\pmb{\theta}$. Note that the denominator in Eq. (\ref{eq:bayes2}) serves to normalize the posterior p.d.f. to unity. The likelihood function, prior, and posterior p.d.f.s all depend on $\pmb{\theta}$, and are related by Bayes' theorem, as usual.

Bayesian statistics does not supply a rule for determining the prior $\pi(\pmb{\theta})$; this reflects the analyst's subjective degree of belief (or state of knowledge) about $\pmb{\theta}$ before the measurement was carried out.

\section{The p.d.f. for more UHOs}

The sum from the next UHO to the optimal truncation, $\Delta_k = \sum_{i=k+1}^{N} c_i \alpha_s^i$, depends on the values of the unknown coefficients $c_{k+1},c_{k+2},\cdots,c_{N}$. It's conditional p.d.f. $f_\Delta(\Delta_k|c_l,\cdots,c_k)$ can be written as
\begin{widetext}
\begin{eqnarray}
\label{eq:UHOspdf1}
f_\Delta(\Delta_k|c_l,\cdots,c_k)=\int \left[\delta\left(\Delta_k-\sum_{n=k+1}^{N}\alpha_s^n c_n\right)\right]f_{rc}(c_{k+1},c_{k+2},\cdots,c_{N}|c_l,\cdots,c_k)\ {\rm d} c_{k+1} {\rm d} c_{k+2} \cdots {\rm d} c_{N}\;,
\end{eqnarray}
\end{widetext}
where $f_{rc}(c_{k+1},c_{k+2},\cdots,c_{N}|c_l,\cdots,c_k)$ is the conditional p.d.f. of $c_{k+1},c_{k+2},\cdots,c_{N}$ given $c_l,\dots,c_k$. This expression is too complicated to be handled analytically. In order to perform a numerical integration of Eq.(\ref{eq:UHOspdf1}), we can rewrite it as
\begin{widetext}
\begin{eqnarray}
\label{eq:UHOspdf2}
f_\Delta(\Delta_k|c_l,\cdots,c_k)=\int \left[\delta\left(\Delta_k-\sum_{n=k+1}^{N}\alpha_s^n c_n\right)\right]
\left[\prod_{n=k+1}^{N} h_0(c_n|\bar c)\right] f_{\bar c}(\bar c| c_l,\dots,c_k)\ {\rm d} {\bar c}\
{\rm d} c_{k+1} {\rm d} c_{k+2} \cdots {\rm d} c_{N}\;.
\end{eqnarray}
\end{widetext}
where $f_{\bar c}({\bar c}|c_l,\dots,c_k)$ is the conditional p.d.f. of ${\bar c}$ given $c_l,\dots,c_k$, which can be obtained by using the Bayes' formula (\ref{eq:bayes}) and taking the limit $\epsilon\to 0$ for the final result,
\begin{eqnarray}
\label{eq:cbarCondpdf}
f_{\bar c}(\bar c|c_l,\dots,c_k)=n_c\frac{\bar{c}_{(k)}^{n_c}}{\bar c^{n_c+1}}\ \theta({\bar c}-\bar{c}_{(k)}).
\end{eqnarray}

\section{A glossary}

{\bf Priori probability}: the probability estimate prior to receiving new information.

{\bf Posterior probability}: the revised probability that takes into account new available information.

{\bf Probability density function}: a non-negative function which describes the distribution of a continuous random variable.

{\bf Random variable}: a variable that takes different real values as a result of the outcomes of a random event or experiment.

{\bf Credibility}: also called ``degree of belief'', ``Bayesian probability'', or ``subjective probability''.

{\bf Credibility measure}: the credibility measure plays a similar role as the probability measure but applies to Bayesian probability.

{\bf UHO}: Unknown Higher-Order

{\bf PMC}: Principle of Maximum Conformality

{\bf PMCs}: PMC single scale-setting approach

{\bf PMCm}: PMC multi scale-setting approach

{\bf RGE}: Renormalization Group Equation

{\bf RGI}: Renormalization Group Invariance

{\bf p.d.f.}: probability density function

{\bf CI}: Credible interval

{\bf DoB}: Degree-of-Belief

{\bf EC}: Exact value


\begin{thebibliography}{100}

\bibitem{Gross:1973id}
 D.~J.~Gross and F.~Wilczek,
 Ultraviolet Behavior of Nonabelian Gauge Theories,
 Phys.\ Rev.\ Lett.\ {\bf 30}, 1343 (1973).

\bibitem{Politzer:1973fx}
 H.~D.~Politzer,
 Reliable Perturbative Results for Strong Interactions?,
 Phys.\ Rev.\ Lett.\ {\bf 30}, 1346 (1973).

\bibitem{Petermann:1953wpa}
 A.~Petermann,
 Normalization of constants in the quanta theory,
 Helv.\ Phys.\ Acta {\bf 26}, 499 (1953).

\bibitem{GellMann:1954fq}
 M.~Gell-Mann and F.~E.~Low,
 Quantum electrodynamics at small distances,
 Phys.\ Rev.\ {\bf 95}, 1300 (1954).

\bibitem{Peterman:1978tb}
 A.~Peterman,
 Renormalization Group and the Deep Structure of the Proton,
 Phys.\ Rept.\ {\bf 53}, 157 (1979).

\bibitem{Callan:1970yg}
 C.~G.~Callan, Jr.,
 Broken scale invariance in scalar field theory,
 Phys.\ Rev.\ D {\bf 2}, 1541 (1970).

\bibitem{Symanzik:1970rt}
 K.~Symanzik,
 Small distance behavior in field theory and power counting,
 Commun.\ Math.\ Phys.\ {\bf 18}, 227 (1970).

\bibitem{Brodsky:2011ig}
 S.~J.~Brodsky and L.~Di Giustino,
 Setting the Renormalization Scale in QCD: The Principle of Maximum Conformality,
 Phys. Rev. D \textbf{86}, 085026 (2012).

\bibitem{Mojaza:2012mf}
 M.~Mojaza, S.~J.~Brodsky, and X.~G.~Wu,
 Systematic All-Orders Method to Eliminate Renormalization-Scale and Scheme Ambiguities in Perturbative QCD,
 Phys.\ Rev.\ Lett.\ {\bf 110}, 192001 (2013).

\bibitem{Brodsky:2013vpa}
 S.~J.~Brodsky, M.~Mojaza, and X.~G.~Wu,
 Systematic Scale-Setting to All Orders: The Principle of Maximum Conformality and Commensurate Scale Relations,
 Phys.\ Rev.\ D {\bf 89}, 014027 (2014).

\bibitem{Brodsky:2011ta}
 S.~J.~Brodsky and X.~G.~Wu,
 Scale Setting Using the Extended Renormalization Group and the Principle of Maximum Conformality: the QCD Coupling Constant at Four Loops,
 Phys.\ Rev.\ D {\bf 85}, 034038 (2012).

\bibitem{Brodsky:2012rj}
 S.~J.~Brodsky and X.~G.~Wu,
 Eliminating the Renormalization Scale Ambiguity for Top-Pair Production Using the Principle of Maximum Conformality,
 Phys.\ Rev.\ Lett.\ {\bf 109}, 042002 (2012).

\bibitem{Brodsky:2012ms}
 S.~J.~Brodsky and X.~G.~Wu,
 Self-Consistency Requirements of the Renormalization Group for Setting the Renormalization Scale,
 Phys.\ Rev.\ D {\bf 86}, 054018 (2012).

\bibitem{Wu:2014iba}
 X.~G.~Wu, Y.~Ma, S.~Q.~Wang, H.~B.~Fu, H.~H.~Ma, S.~J.~Brodsky and M.~Mojaza,
 Renormalization Group Invariance and Optimal QCD Renormalization Scale-Setting,
 Rept.\ Prog.\ Phys.\ {\bf 78}, 126201 (2015).

\bibitem{Wu:2013ei}
 X.~G.~Wu, S.~J.~Brodsky and M.~Mojaza,
 The Renormalization Scale-Setting Problem in QCD,
 Prog.\ Part.\ Nucl.\ Phys.\ {\bf 72}, 44 (2013).

\bibitem{Gross:1973ju}
 D.~J.~Gross and F.~Wilczek,
 Asymptotically Free Gauge Theories - I,
 Phys.\ Rev.\ D {\bf 8}, 3633 (1973).

\bibitem{Politzer:1974fr}
 H.~D.~Politzer,
 Asymptotic Freedom: An Approach to Strong Interactions,
 Phys.\ Rept.\ {\bf 14}, 129 (1974).

\bibitem{Caswell:1974gg}
 W.~E.~Caswell,
 Asymptotic Behavior of Nonabelian Gauge Theories to Two Loop Order,
 Phys.\ Rev.\ Lett.\ {\bf 33}, 244 (1974).

\bibitem{Tarasov:1980au}
 O.~V.~Tarasov, A.~A.~Vladimirov and A.~Y.~Zharkov,
 The Gell-Mann-Low Function of QCD in the Three Loop Approximation,
 Phys.\ Lett.\ B {\bf 93}, 429 (1980).

\bibitem{Larin:1993tp}
 S.~A.~Larin and J.~A.~M.~Vermaseren,
 The Three loop QCD Beta function and anomalous dimensions,
 Phys.\ Lett.\ B {\bf 303}, 334 (1993).

\bibitem{vanRitbergen:1997va}
 T.~van Ritbergen, J.~A.~M.~Vermaseren and S.~A.~Larin,
 The Four loop beta function in quantum chromodynamics,
 Phys.\ Lett.\ B {\bf 400}, 379 (1997).

\bibitem{Chetyrkin:2004mf}
 K.~G.~Chetyrkin,
 Four-loop renormalization of QCD: Full set of renormalization constants and anomalous dimensions,
 Nucl.\ Phys.\ B {\bf 710}, 499 (2005).

\bibitem{Czakon:2004bu}
 M.~Czakon,
 The Four-loop QCD beta-function and anomalous dimensions,
 Nucl.\ Phys.\ B {\bf 710}, 485 (2005).

\bibitem{Baikov:2016tgj}
 P.~A.~Baikov, K.~G.~Chetyrkin and J.~H.~Kuhn,
 Five-Loop Running of the QCD coupling constant,
 Phys.\ Rev.\ Lett.\ {\bf 118}, 082002 (2017).

\bibitem{Brodsky:1982gc}
 S.~J.~Brodsky, G.~P.~Lepage and P.~B.~Mackenzie,
 On the Elimination of Scale Ambiguities in Perturbative Quantum Chromodynamics,
 Phys.\ Rev.\ D {\bf 28}, 228 (1983).

\bibitem{Brodsky:1997jk}
 S.~J.~Brodsky and P.~Huet,
 Aspects of SU(N(c)) gauge theories in the limit of small number of colors,
 Phys.\ Lett.\ B {\bf 417}, 145 (1998).

\bibitem{Brodsky:1994eh}
 S.~J.~Brodsky and H.~J.~Lu,
 Commensurate scale relations in quantum chromodynamics,
 Phys.\ Rev.\ D {\bf 51}, 3652 (1995).

\bibitem{Huang:2020gic}
 X.~D.~Huang, X.~G.~Wu, Q.~Yu, X.~C.~Zheng, J.~Zeng and J.~M.~Shen,
 Generalized Crewther relation and a novel demonstration of the scheme independence of commensurate scale relations up to all orders,
 Chin.\ Phys.\ C {\bf 45}, 103104 (2021).

\bibitem{Beneke:1998ui}
 M.~Beneke,
 Renormalons,
 Phys.\ Rept.\ {\bf 317}, 1 (1999).

\bibitem{Beneke:1994qe}
 M.~Beneke and V.~M.~Braun,
 Naive nonAbelianization and resummation of fermion bubble chains,
 Phys.\ Lett.\ B {\bf 348}, 513 (1995).

\bibitem{Neubert:1994vb}
 M.~Neubert,
 Scale setting in QCD and the momentum flow in Feynman diagrams,
 Phys.\ Rev.\ D {\bf 51}, 5924 (1995).

\bibitem{Cacciari:2011ze}
M.~Cacciari and N.~Houdeau,
Meaningful characterization of perturbative theoretical uncertainties,
JHEP \textbf{09}, 039 (2011).

\bibitem{Bagnaschi:2014wea}
E.~Bagnaschi, M.~Cacciari, A.~Guffanti and L.~Jenniches,
An extensive survey of the estimation of uncertainties from missing higher orders in perturbative calculations,
JHEP \textbf{02}, 133 (2015).

\bibitem{Bonvini:2020xeo}
M.~Bonvini,
Probabilistic definition of the perturbative theoretical uncertainty from missing higher orders,
Eur. Phys. J. C \textbf{80}, 989 (2020).

\bibitem{Duhr:2021mfd}
C.~Duhr, A.~Huss, A.~Mazeliauskas and R.~Szafron,
An analysis of Bayesian estimates for missing higher orders in perturbative calculations,
JHEP \textbf{09}, 122 (2021).

\bibitem{Workman:2022ynf}
R.~L.~Workman [Particle Data Group],
Review of Particle Physics,
PTEP \textbf{2022}, 083C01 (2022).

\bibitem{Dyson:1952tj}
F.~J.~Dyson,
Divergence of perturbation theory in quantum electrodynamics,
Phys. Rev. \textbf{85}, 631-632 (1952).

\bibitem{tHooft:1977xjm}
G.~'t Hooft,
Can We Make Sense Out of Quantum Chromodynamics?,
Subnucl. Ser. \textbf{15}, 943 (1979).

\bibitem{BO1}
 D.~R.~Jones, M.~Schonlau and W.~J.~Welch,
 Efficient Global Optimization of Expensive Black-Box Functions,
 J.\ Global Optim.\ {\bf 13}, 455 (1998).

\bibitem{BO2}
 E.~Brochu, V.~M.~Cora, and N.~de Freitas,
 A Tutorial on Bayesian Optimization of Expensive Cost Functions, with Application to Active User Modeling and Hierarchical Reinforcement Learning,
 arXiv:1012.2599.

\bibitem{Zheng:2013uja}
 X.~C.~Zheng, X.~G.~Wu, S.~Q.~Wang, J.~M.~Shen, and Q.~L.~Zhang,
 Reanalysis of the BFKL Pomeron at the next-to-leading logarithmic accuracy,
 J. High Energy Phys. {\bf 10}, 117 (2013).

\bibitem{Wu:2019mky}
 X.~G.~Wu, J.~M.~Shen, B.~L.~Du, X.~D.~Huang, S.~Q.~Wang, and S.~J.~Brodsky,
 The QCD Renormalization Group Equation and the Elimination of Fixed-Order Scheme-and-Scale Ambiguities Using the Principle of Maximum Conformality,
 Prog.\ Part.\ Nucl.\ Phys.\ {\bf 108}, 103706 (2019).

\bibitem{Huang:2021hzr}
X.~D.~Huang, J.~Yan, H.~H.~Ma, L.~Di Giustino, J.~M.~Shen, X.~G.~Wu and S.~J.~Brodsky,
Detailed comparison of renormalization scale-setting procedures based on the principle of maximum conformality,
Nucl. Phys. B \textbf{989}, 116150 (2023).

\bibitem{Shen:2017pdu}
 J.~M.~Shen, X.~G.~Wu, B.~L.~Du and S.~J.~Brodsky,
 Novel All-Orders Single-Scale Approach to QCD Renormalization Scale-Setting,
 Phys.\ Rev.\ D {\bf 95}, 094006 (2017).

\bibitem{Wu:2018cmb}
 X.~G.~Wu, J.~M.~Shen, B.~L.~Du, and S.~J.~Brodsky,
 Novel demonstration of the renormalization group invariance of the fixed-order predictions using the principle of maximum conformality and the $C$-scheme coupling,
 Phys.\ Rev.\ D {\bf 97}, 094030 (2018).

\bibitem{Yan:2022foz}
J.~Yan, Z.~F.~Wu, J.~M.~Shen and X.~G.~Wu,
``Precise perturbative predictions from fixed-order calculations,''
J. Phys. G \textbf{50}, 045001 (2023).

\bibitem{DiGiustino:2020fbk}
L.~Di Giustino, S.~J.~Brodsky, S.~Q.~Wang and X.~G.~Wu,
``Infinite-order scale-setting using the principle of maximum conformality: A remarkably efficient method for eliminating renormalization scale ambiguities for perturbative QCD,''
Phys. Rev. D \textbf{102}, 014015 (2020).

\bibitem{Bi:2015wea}
 H.~Y.~Bi, X.~G.~Wu, Y.~Ma, H.~H.~Ma, S.~J.~Brodsky and M.~Mojaza,
 Degeneracy Relations in QCD and the Equivalence of Two Systematic All-Orders Methods for Setting the Renormalization Scale,
 Phys.\ Lett.\ B {\bf 748}, 13 (2015).

\bibitem{Baikov:2008jh}
 P.~A.~Baikov, K.~G.~Chetyrkin and J.~H.~Kuhn,
 Order $\alpha^4(s)$ QCD Corrections to $Z$ and tau Decays,
 Phys.\ Rev.\ Lett.\ {\bf 101}, 012002 (2008).

\bibitem{Baikov:2010je}
 P.~A.~Baikov, K.~G.~Chetyrkin and J.~H.~Kuhn,
 Adler Function, Bjorken Sum Rule, and the Crewther Relation to Order $\alpha_s^4$ in a General Gauge Theory,
 Phys.\ Rev.\ Lett.\ {\bf 104}, 132004 (2010).

\bibitem{Baikov:2012zn}
 P.~A.~Baikov, K.~G.~Chetyrkin, J.~H.~Kuhn and J.~Rittinger,
 Adler Function, Sum Rules and Crewther Relation of Order ${\cal O}(\alpha_s^4)$: the Singlet Case,
 Phys.\ Lett.\ B {\bf 714}, 62 (2012).

\bibitem{Baikov:2012zm}
 P.~A.~Baikov, K.~G.~Chetyrkin, J.~H.~Kuhn and J.~Rittinger,
 Vector Correlator in Massless QCD at Order ${\cal O}(\alpha_s^4)$ and the QED beta-function at Five Loop,
 JHEP {\bf 1207}, 017 (2012).

\bibitem{Marshall:1988ri}
 R.~Marshall,
 A Determination of the Strong Coupling Constant $\alpha^- s$ From $e^+ e^-$ Total Cross-section Data,
 Z.\ Phys.\ C {\bf 43}, 595 (1989).

\bibitem{Lam:1977cu}
 C.~S.~Lam and T.~-M.~Yan,
 Decays of Heavy Lepton and Intermediate Weak Boson in Quantum Chromodynamics,
 Phys.\ Rev.\ D {\bf 16}, 703 (1977).

\bibitem{Chetyrkin:2000yt}
 K.~G.~Chetyrkin, J.~H.~Kuhn and M.~Steinhauser,
 RunDec: A Mathematica package for running and decoupling of the strong coupling and quark masses,
 Comput.\ Phys.\ Commun.\ {\bf 133}, 43 (2000).

\bibitem{Herren:2017osy}
 F.~Herren and M.~Steinhauser,
 Version 3 of RunDec and CRunDec,
 Comput.\ Phys.\ Commun.\ {\bf 224}, 333 (2018).

\bibitem{Beneke:2008ad}
M.~Beneke and M.~Jamin,
``alpha(s) and the tau hadronic width: fixed-order, contour-improved and higher-order perturbation theory,''
JHEP \textbf{09}, 044 (2008).

\bibitem{Boito:2016pwf}
D.~Boito, M.~Jamin and R.~Miravitllas,
``Scheme Variations of the QCD Coupling and Hadronic \ensuremath{\tau} Decays,''
Phys. Rev. Lett. \textbf{117}, 152001 (2016).

\bibitem{Boito:2018rwt}
D.~Boito, P.~Masjuan and F.~Oliani,
``Higher-order QCD corrections to hadronic $\tau$ decays from Pad\'e approximants,''
JHEP \textbf{08}, 075 (2018).

\bibitem{Caprini:2018agy}
I.~Caprini,
``Renormalization-scheme variation of a QCD perturbation expansion with tamed large-order behavior,''
Phys. Rev. D \textbf{98}, 056016 (2018).

\bibitem{Kolmogorov:1933}
A.~N.~Kolmogorov, Grundbegriffe der Wahrscheinlichkeitsrechnung (Springer, Berlin, 1933).

\bibitem{Kolmogorov:1956}
Foundations of the Theory of Probability, 2nd edn. (Chelsea, New York 1956).

\end{thebibliography}
\end{document}